\documentclass[journal,10pt]{IEEEtran}
\usepackage{amsmath,amsfonts}
\usepackage[ruled, linesnumbered]{algorithm2e}
\usepackage{array}
\usepackage[caption=false,font=normalsize,labelfont=sf,textfont=sf]{subfig}
\usepackage{textcomp}
\usepackage{stfloats}
\usepackage{url}
\usepackage{verbatim}
\usepackage{graphicx}
\usepackage{cite}
\usepackage{hyperref}
\usepackage{cleveref}
\usepackage{subfig}
\usepackage{flushend}
\usepackage{bbding}
\begin{document}

\title{Fast and Arbitrary Beam Pattern Design for RIS-Assisted Terahertz Wireless Communication}

\author{Jian Dang,~\IEEEmembership{Senior Member,~IEEE},
Zaichen Zhang\Envelope,~\IEEEmembership{Senior Member,~IEEE},
Yewei Li,
Liang Wu,~\IEEEmembership{Member,~IEEE},
Bingcheng Zhu,~\IEEEmembership{Member,~IEEE},
and
Lei Wang,~\IEEEmembership{Member,~IEEE}

\thanks{J. Dang, Z. Zhang, Y. Li, L. Wu, B. Zhu, and L. Wang are with the National Mobile Communications Research Laboratory, Southeast University, Nanjing 210096, China. J. Dang, Z. Zhang, L. Wu, B. Zhu, and L. Wang are also with the Purple Mountain Laboratory, Nanjing 211111, China. E-mails: \{dangjian,zczhang,220190835,wuliang,zbc,wang\_lei\_seu\}@seu.edu.cn.}
\thanks{Manuscript received XXX, XX, 2015; revised XXX, XX, 2015.}}

\markboth{IEEE Transactions on Vehicular Technology,~Vol.~XX, No.~XX, XXX~2015}
{}
\maketitle

\begin{abstract}
 Reconfigurable intelligent surface (RIS) can assist terahertz wireless communication to restore the fragile line-of-sight links and facilitate beam steering. Arbitrary reflection beam patterns are desired to meet diverse requirements in different applications. This paper establishes relationship between RIS beam pattern design with two-dimensional finite impulse response filter design and proposes a fast non-iterative algorithm to solve the problem. Simulations show that the proposed method outperforms baseline method. Hence, it represents a promising solution for fast and arbitrary beam pattern design in RIS-assisted terahertz wireless communication. 
\end{abstract}


\IEEEpeerreviewmaketitle

\section{Introduction}\label{sec1}
Terahertz frequency entitles excessive spectrum bandwidth and sub-millimeter wavelength and plays important roles in wireless communications, sensing, medical imaging, and other relevant applications \cite{ChMZ19,RaXK19,CSSB22}. In terahertz wireless communication, the link is line-of-sight (LOS) dominant which may be blocked by random obstacles. To alleviate this adverse effect, reconfigurable intelligent surface (RIS) were introduced in terahertz wireless communication \cite{BoAl21,CNHT21,PWPZ22}. Through beam reflection, an indirect LOS link can be restored. In this process, the reflection coefficients of the RIS units must be carefully designed to guarantee a desired reflection behavior. 

In general, complicated optimization with accurate channel state information (CSI) is required to achieve the beamforming goal. This may hinder the applicability of RIS in wireless communication with mobile users. On the other hand, existing works mainly focus on maximizing the received power for given users. While in mobile communications, users may move around and information broadcasting with wide beam is more suitable in this case. In other words, arbitrary beamforming without explicit CSI is a desired functionality of RIS-assisted wireless communication. While this may be hard in lower frequency band, it is possible in terahertz communication since the link is LOS dominant, as will be shown in this work. With this functionality, we can realize many applications such as broadcasting, localization and mobile communication in a flexible and convenient way. 

However, realizing the functionality purely based on finite number of reflection units is not an easy task. To the best of our knowledge, only limited works had touched this topic. In \cite{HeXZ21}, RIS assisted broadband coverage for millimeter wave communication was formulated into a manifold optimization problem. Massive number of antennas at the base station (BS) and uniform linear array arrangement of RIS are key assumptions in the derivation. Thus only 2-dimensional (2-D) beam can be generated through complex optimization. In addition, only rectangular beam was tested and whether the beams with arbitrary cross shapes can be supported remained unclear. \cite{WGGR21} studied RIS beam design for terahertz wireless communication. Arbitrary cross-shaped beam patterns were supported through a non-iterative design procedure. Nonetheless, as the equivalent transform domain response were designated directly without joint optimization, the overall performance could be poor with limited number of RIS units.


In this work, motivated by \cite{WGGR21}, we develop a different approach to obtain the reflection coefficients. The core idea is to view the beam pattern as a 2-D Fourier transform. With this regard, we establish the equivalence between arbitrary RIS beam pattern design and 2-D finite impulse response (FIR) filter design through a series of transformations. By leveraging a novel existing filter design method with conjugate filter coefficients, closed-form expression for the design problem is obtained. The contribution of this paper is summarized as follows:
(1) We propose an arbitrary beam pattern design method for RIS-assisted terahertz wireless communication. The reflection coefficients have closed-form and the performance is superior to existing solution.
(2) As the closed-form expression is still computationally demanding, we further propose a simplified solution using 2-D inverse fast Fourier transform (IFFT).

The remainder of this paper is organized as follows. Section \ref{sec:system model} provides system model and problem formulation. Section \ref{sec:design} analyse the problem and proposes a closed-form solution. Some practical issues are also discussed. Section \ref{sec:sim} conducts simulations to verify the analysis. Section \ref{sec:conclusion} concludes the paper.

\section{System Model}\label{sec:system model}
\subsection{System Description}
\begin{figure}[htb]
\centering
\includegraphics[width=0.49\textwidth]{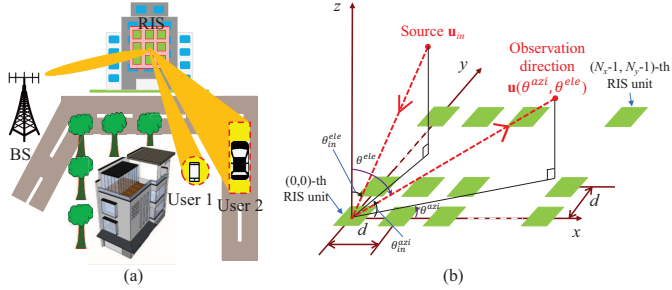}
\caption{System model of RIS-assisted terahertz communication. (a) communication scenario; (b) geometry arrangement of RIS.}
\label{fig:system model}
\end{figure}

Fig. \ref{fig:system model}(a) depicts a typical deployment scenario of RIS-assisted terahertz communication system. Due to blockage, some users in a certain area cannot establish direct LOS communication links with the BS. This is particularly undesirable in terahertz band since the path loss with shadowing is very severe. To recover the communication links, a RIS is deployed to reflect the electromagnetic wave in a controllable manner for the users in the blind area. LOS paths are assumed for the BS-RIS and RIS-users links. For the BS, we assume it consists of multiple antennas such that a beam with sufficient energy is shaped and targeted toward the RIS.

Through careful designing of its reflection coefficients, the RIS receives the incident beam from the BS and scatters the beam into multiple beams to serve multiple users. Each reflected beam may have different cross shapes and power allocations. As an example, Fig. \ref{fig:system model}(a) shows two users served by a single RIS. User 1 is static, so the RIS allocates it a narrower beam with a circle shape whose diameter is related to the positioning accuracy of user 1. User 2 is moving, so RIS generates a wider beam with a rectangular shape for steady coverage. Note that the two beams are generated simultaneously by properly setting the reflection coefficients, which is the focus of this paper.

Fig. \ref{fig:system model}(b) provides a closer look on the reflection behaviour of the RIS in a three-dimensional Cartesian coordinates system. The RIS has $N_x\times N_y$ reflection units with $d$ as the space between two adjacent units. The RIS is placed in the $x$-$y$ plane and the center of its $(0,0)$-th unit is set as the origin. Far field effect is considered in this paper so the incident waves impinging the RIS are plane waves. Assume the direction of the incident wave is determined by the horizontal (azimuth) angle $\theta_{in}^{azi}\in[0,2\pi)$ and the elevation angle $\theta_{in}^{ele}\in[0,\frac{\pi}{2})$, where the definitions of $\theta_{in}^{azi}$ and $\theta_{in}^{ele}$ are shown in Fig. \ref{fig:system model}(b). They are fixed and known a priori. Thus, the unit directional vector of the incident wave can be denoted by
\begin{align}\label{eq:uin}
   \mathbf{u}_{in}=[\cos\theta_{in}^{azi}\sin\theta_{in}^{ele},\sin\theta_{in}^{azi}\sin\theta_{in}^{ele},-\cos\theta_{in}^{ele}]^T. 
\end{align}
For the $(n_x,n_y)$-th reflection unit, the phase difference of its received signal with respect to that of the reference unit, i.e., the $(0,0)$-th unit, is determined by the difference of the transmission distance from the BS, which is approximately equals to $\mathbf{u}_{in}^T\mathbf{r}(n_x,n_y)$, where $\mathbf{r}(n_x,n_y)=[n_xd,n_yd,0]^T$ is the directional vector from the origin to the center of the $(n_x,n_y)$-th unit. Therefore, the phase difference is given by $e^{j\frac{2\pi}{\lambda}\mathbf{u}_{in}^T\mathbf{r}(n_x,n_y)}$. 

Define $\theta^{azi}$ and $\theta^{ele}$ as the azimuth and the elevation angles of an observation direction of the reflected signal which are depicted in Fig. \ref{fig:system model}(b). Here, $\theta^{azi}$ and $\theta^{ele}$ are treated as two variables rather than two constants. The unit directional vector of the reflected wave as a function of  $\theta^{azi}$ and $\theta^{ele}$ can be denoted by
\begin{align}
   \mathbf{u}(\theta^{azi}\!,\theta^{ele} )\!=\![\cos\theta^{azi}\!\sin\theta^{ele}\!,\!\sin\theta^{azi}\sin\theta^{ele}\!,\!\cos\theta^{ele}]^T. 
\end{align}
With a similar analysis on the phase difference for the input signal, the phase difference between the reflected signals from the $(n_x,n_y)$-th unit and the reference unit, perceived by the observer located far away in the direction of $\mathbf{u}(\theta^{azi},\theta^{ele})$, is given by $e^{-j\frac{2\pi}{\lambda}\mathbf{u}^T(\theta^{azi},\theta^{ele} )\mathbf{r}(n_x,n_y)}$.

Denote the reflection coefficient of the $(n_x,n_y)$-th unit by $v(n_x,n_y)$ where $v(n_x,n_y)$ is a complex number. Given any observation direction  $\mathbf{u}(\theta^{azi},\theta^{ele})$, the aggregate response is the sum of the reflected signals from all the RIS units with perspective phases:
\begin{align}\label{eq:g theta}
\nonumber    g(\theta^{azi},\theta^{ele})
     =\sum_{n_x=0}^{N_x-1}\sum_{n_y=0}^{N_y-1} e^{j\frac{2\pi}{\lambda}\mathbf{u}_{in}^T\mathbf{r}(n_x,n_y)} v(n_x,n_y)\\ e^{-j\frac{2\pi}{\lambda}\mathbf{u}^T(\theta^{azi},\theta^{ele} )\mathbf{r}(n_x,n_y)}.
\end{align}
$g(\theta^{azi},\theta^{ele})$ is a function of $\theta^{azi}$ and $\theta^{ele}$ characterizing the response of the reflected signal among the three-dimensional space. It should be noted that $g(\theta^{azi},\theta^{ele})$ is a relative quantity without considering the path loss effect. However, the difference in path loss for different observation directions can be compensated by designing the power distribution of $g(\theta^{azi},\theta^{ele})$, which will be elaborated later. $g(\theta^{azi},\theta^{ele})$ is thus described as the (reflected) beam pattern \cite{WGGR21}.
\subsection{Problem Formulation}
The problem is to find an optimal set of $\mathbf{v}=\{v(n_x,n_y),n_x=0,1,\cdots,N_x-1,n_y=0,1,\cdots,N_y-1\}$, such that the resultant beam pattern $g(\theta^{azi},\theta^{ele})$ is as close as to a desired beam pattern $\hat{g}(\theta^{azi},\theta^{ele})$. The desired beam pattern is given a priori according to the specific applications, such as shown in  Fig. \ref{fig:system model}(a). In practice, the phase response of the desired beam pattern is hard to manipulate, so we only focus on the magnitude response. In summary, the problem is formally given by (\textbf{P1}):
\begin{align}
\nonumber   (\textbf{P1}) \indent \mathbf{v}^{opt}=\arg \min_{\{\mathbf{v}\}} \int_{0}^{\frac{\pi}{2}} \int_{0}^{2\pi} [|g(\theta^{azi},\theta^{ele})|\\
   -|\hat{g}(\theta^{azi},\theta^{ele})| ]^2 \text{d}\theta^{azi}\text{d}\theta^{ele}.
\end{align}
Note that we do not impose any constraint on $v(n_x,n_y)$ at this stage. This is also the case in \cite{WGGR21}. In practice, hardware constraints such as limited resolutions of the amplitude and phase of the reflection coefficients will be imposed through quantization on $\mathbf{v}^{opt}$, which is validated in simulations. 

\section{Beam Pattern Design}\label{sec:design}
The response form in \eqref{eq:g theta} shares similarity with the response of 2-D FIR filters. In this paper, we show that through a series of transformations, the original problem of (P1) is equivalent to the problem of designing 2-D FIR filters, which has been studied for a long time and exists some simple solutions. To elaborate the transformations, we first briefly review the designing of 2-D FIR filters.
\subsection{2-D FIR Filter Design with Closed-Form Expression}
2-D FIR filter design is a well studied topic in signal processing. Among various works, we are interested in how to design a filter with arbitrary magnitude response. In the total squared-error (TSE) sense, there exists a closed-form optimal solution for this problem which we will review briefly here \cite{ZhAS97}. Considering a 2-D FIR filter $h(m,n)$, where $m=0,1,\cdots,L_1-1$, $n=0,1,\cdots,L_2-1$, its frequency domain response is given by
\begin{align}\label{eq:He}
    H(e^{j\omega_1},e^{j\omega_2})=\sum_{m=0}^{L_1-1}\sum_{n=0}^{L_2-1} h(m,n)e^{-j(m\omega_1 +n\omega_2)}.
\end{align}
The key of the approach in \cite{ZhAS97} is to impose Hermitian symmetry on the coefficients, i.e., 
\begin{align}\label{eq:conj}
    h(m,n)=h^*(L_1-1-m,L_2-1-n),
\end{align}
such that its magnitude response can be separated from its phase response, as shown in the following:
\begin{align}
    H(e^{j\omega_1},e^{j\omega_2})=e^{-j(\frac{L_1-1}{2}\omega_1+\frac{L_2-1}{2}\omega_2)}H(\omega_1,\omega_2),
\end{align}
where the complex exponential part denotes the linear-phase response of the filter, and the purely real function $H(\omega_1,\omega_2)$ is the magnitude response of the filter. Assume the desired magnitude response is denoted by $\hat{H}(\omega_1,\omega_2)$. It is uniformly sampled over the $(\omega_1,\omega_2)$-plane by using an $M_1\times M_2$ rectangular grid, producing the samples as $\hat{H}(\omega_{1,k},\omega_{2,l})$, where the points on the grid is defined as
\begin{align}\label{eq:omega12}
    \omega_{1,k}=\frac{2\pi k}{M_1}-\pi, \indent \omega_{2,l}=\frac{2\pi l}{M_2}-\pi, 
\end{align}
for $k=0,1,\cdots,M_1-1$, and $l=0,1,\cdots,M_2-1$. TSE is used to evaluate the accuracy of the designed filter, which is given by
\begin{align}\label{eq:E}
TSE=\sum_{k=0}^{M_1-1}\sum_{l=0}^{M_2-1}\left[H(\omega_{1,k},\omega_{2,l})-\hat{H}(\omega_{1,k},\omega_{2,l})\right]^2.
\end{align}
Through matrix manipulations, it is proved that the desired filter coefficients which minimize the TSE have closed-form expression given by:
\begin{align}\label{eq:h}
\nonumber    h(m,n)=\frac{1}{M_1M_2}\sum_{k=0}^{M_1-1}\sum_{l=0}^{M_2-1}\hat{H}(\omega_{1,k},\omega_{2,l})\\e^{j[(m-\frac{L_1-1}{2})\omega_{1,k}+(n-\frac{L_2-1}{2})\omega_{2,l}]},
\end{align}
for $m=0,1,\cdots,\frac{L_1}{2}-1$, and $n=0,1,\cdots,L_2-1$, when $L_1$ and $L_2$ are both even numbers\footnote{Other combinations of $L_1$ and $L_2$ also have similar solutions. This paper assumes $L_1$ and $L_2$ are both even numbers.}. The other half of the coefficients can be obtained through \eqref{eq:conj}. 

\subsection{Transform Domain Representation}
In this subsection, we show that (P1) can be solved by leveraging the aforementioned 2-D FIR filter design approach and even simpler expression with fast algorithm can be obtained as well.

Firstly, we define
\begin{align}\label{eq:hnxny}
\nonumber    &h(n_x,n_y)=e^{j\frac{2\pi}{\lambda}\mathbf{u}_{in}^T\mathbf{r}(n_x,n_y)} v(n_x,n_y)\\
    &=v(n_x,n_y) e^{-j(n_x\frac{2\pi d}{\lambda}\cos\theta_{in}^{azi}\sin\theta_{in}^{ele}+n_y\sin\theta_{in}^{azi}\sin\theta_{in}^{ele})},
\end{align}
and
\begin{align}
    \omega_1(\theta^{azi},\theta^{ele})=\frac{2\pi d}{\lambda}\cos\theta^{azi}\sin\theta^{ele},\\
\omega_2(\theta^{azi},\theta^{ele})=\frac{2\pi d}{\lambda}\sin\theta^{azi}\sin\theta^{ele}.
\end{align}
Then the response of \eqref{eq:g theta} can be rewritten as
\begin{align}\label{eq:g transform}
g(\theta^{azi},\theta^{ele})
=\sum_{n_x=0}^{N_x-1}\sum_{n_y=0}^{N_y-1} h(n_x,n_y)e^{-j\left[n_x\omega_1+n_y\omega_2\right]}.
\end{align}
For notational simplicity, we use $\omega_1$ and $\omega_2$ to represent $\omega_1(\theta^{azi},\theta^{ele})$ and $\omega_2(\theta^{azi},\theta^{ele})$, respectively. The form of \eqref{eq:g transform} is very similar to the response of an FIR filter $h(n_x,n_y)$ defined in \eqref{eq:He}. However, $\omega_1$ and $\omega_2$ are not independent variables and certain preconditions must be satisfied to correctly use the approach in \cite{ZhAS97}.

There is a one-to-one mapping from a pair of ($\theta^{azi}, \theta^{ele}$) to a pair of ($\omega_1,\omega_2$). However, the opposite does not hold. Therefore, for any ($\theta^{azi}, \theta^{ele}$) where $|\hat{g}(\theta^{azi}, \theta^{ele})|\neq 0$, the corresponding ($\omega_1,\omega_2$) must satisfy $-\pi \leq \omega_1<\pi$ and $-\pi \leq \omega_2<\pi$ for meaningful interpretation of \eqref{eq:g transform} as filter response. Thus we must have \begin{align}\label{eq:d}
d\leq \frac{\lambda}{2}\min \left\{\frac{1}{|\cos\theta^{azi}\sin\theta^{ele}|},\frac{1}{|\sin\theta^{azi}\sin\theta^{ele}|}\right\},
\end{align}
for those ($\theta^{azi}, \theta^{ele}$) where $|\hat{g}(\theta^{azi}, \theta^{ele})|\neq 0$. An upper bound of $d$ that satisfy the condition of \eqref{eq:d} for all ($\theta^{azi}, \theta^{ele}$) is given by $d=\frac{\lambda}{2}$, which will be assumed in the following of this paper.

\begin{figure}[htb]
    \centering
    \includegraphics[width=1\linewidth,draft=false]{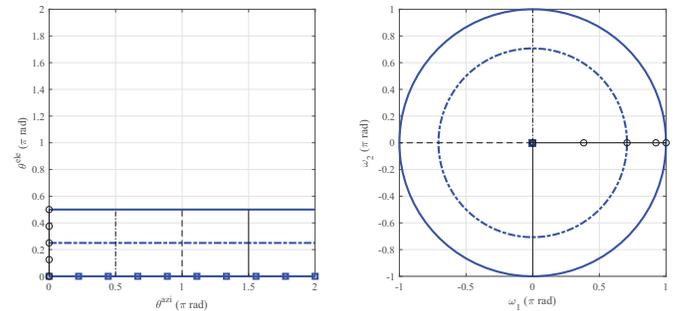}
    \caption{The mapping of $(\theta^{azi},\theta^{ele})$-domain to $( \omega_1, \omega_2)$-domain.}
    \label{fig:transform domain}
\end{figure}

Fig. \ref{fig:transform domain} shows the mapping of $(\theta^{azi},\theta^{ele})$-domain to $( \omega_1, \omega_2)$-domain for $d=\frac{\lambda}{2}$. It can be seen that for any $(\theta^{azi},\theta^{ele})$, the corresponding $(\omega_1, \omega_2)$ are within the range of $[-\pi,\pi)\times[-\pi,\pi)$. On the other hand, the corner parts the $(\omega_1, \omega_2)$-domain do not correspond to any $(\theta^{azi},\theta^{ele})$, which will be set to zero in the designing procedure.

Next, we define $\hat{H}(\omega_{1},\omega_{2})$ and $H(\omega_{1},\omega_{2})$ as the desired and designed magnitude response of $h(n_x,n_y)$, respectively. We samples the $(\omega_1,\omega_2)$-plane with a grid of $M_1\times M_2$ points defined the same as \eqref{eq:omega12}. By denoting \begin{align}
    H(\omega_{1,k},\omega_{2,l})=\hat{H}(\omega_{1,k},\omega_{2,l})=0,
\end{align}
when $\sqrt{\omega_{1,k}^2+\omega_{2,l}^2}>\pi$ and otherwise,
\begin{align}
  H(\omega_{1,k},\omega_{2,l})= |{g}(\theta^{azi}_{k,l},\theta^{ele}_{k,l})|,\\
  \hat{H}(\omega_{1,k},\omega_{2,l})=|\hat{g}(\theta^{azi}_{k,l},\theta^{ele}_{k,l})|,
\end{align} 
where $\theta^{azi}_{k,l},\theta^{ele}_{k,l}$ satisfy 
\begin{align}
    \omega_{1,k}=\pi\cos\theta^{azi}_{k,l}\sin\theta^{ele}_{k,l},\indent
\omega_{2,l}=\pi\sin\theta^{azi}_{k,l}\sin\theta^{ele}_{k,l},
\end{align}
the original problem (P1) can be discretized to the 2-D FIR filter design problem described in \eqref{eq:E} and solved by \eqref{eq:h}. Finally, the reflection coefficients $v(n_x,n_y)$ can be obtained according to \eqref{eq:hnxny}:
\begin{align}
    v(n_x,n_y)=h(n_x,n_y)e^{-j\frac{2\pi}{\lambda}\mathbf{u}_{in}^T\mathbf{r}(n_x,n_y)}.
\end{align}


\subsection{Extension to Multiple Incident Wave Directions}
With the above method, the designing procedure can be extended to multiple incident wave directions. This is useful when a RIS serves multiple BSs or a single BS with distributed antennas. The signal model is almost the same as \eqref{eq:g transform}, except that $h(n_x,n_y)$ is redefined by
\begin{align}
 h(n_x,n_y)=\sum_{i=1}^I e^{j\frac{2\pi}{\lambda}(\mathbf{u}_{in}^{(i)})^T\mathbf{r}(n_x,n_y)} v(n_x,n_y),
\end{align}
where $\mathbf{u}_{in}^{(i)}$ denotes the unit directional vector of the $i$-th incident wave, $i=1,2,\cdots,I$, given in the form of \eqref{eq:uin}. After obtaining $h(n_x,n_y)$, $v(n_x,n_y)$ can be calculated by
\begin{align}\label{eq:vfinal}
 v(n_x,n_y)=\frac{h(n_x,n_y)}{\sum_{i=1}^I e^{j\frac{2\pi}{\lambda}(\mathbf{u}_{in}^{(i)})^T\mathbf{r}(n_x,n_y)}}.
\end{align}

\subsection{Fast Implementation}
While the proposed design has closed-form solution, its computational complexity is still undesirably high. According to \eqref{eq:h}, the computation of all the coefficients has complexity of order $\mathcal{O}(N_xN_yM_1M_2)$, which is very high as there can be a large number of reflection units in the high frequency band. To alleviate this computational burden, we notice that \eqref{eq:h} may be implemented by leveraging 2-D IFFT. To elaborate this, it is noted that the 2-D IFFT of a matrix $\mathbf{X}\in \mathcal{C}^{M\times N}$ is given by
\begin{align}\label{eq:x}
    x(p,q)=\frac{1}{MN}\sum_{m=0}^{M-1}\sum_{n=0}^{N-1} X(m,n)e^{j\frac{2\pi}{M}mp}e^{j\frac{2\pi}{N}nq},
\end{align}
for $p=0,1,\cdots,M-1$, and $q=0,1,\cdots,N-1$, which can be simply computed by sequentially using one-dimensional IFFT twice with size $M$ and $N$, respectively. With this definition, we rewrite \eqref{eq:h} as
\begin{align}\label{eq:h2}
\nonumber    h(m,n)=\frac{1}{M_1M_2}e^{-j(m+n)\pi} \sum_{k=0}^{M_1-1}\sum_{l=0}^{M_2-1}\hat{H}(\omega_{1,k},\omega_{2,l})\\e^{-j(\frac{L_1-1}{2}\omega_{1,k}+\frac{L_2-1}{2}\omega_{2,l})}e^{j(\frac{2\pi}{M_1}km+\frac{2\pi}{M_2}ln)}.
\end{align}
By defining 
\begin{align}\label{eq:tildeH}
\tilde{H}(\omega_{1,k},\omega_{2,l})=\hat{H}(\omega_{1,k},\omega_{2,l})e^{-j(\frac{L_1-1}{2}\omega_{1,k}+\frac{L_2-1}{2}\omega_{2,l})},
\end{align}
its 2-D IFFT is just given by
\begin{align}\label{eq:tildeh}
\tilde{h}(m,n)=\frac{1}{M_1M_2 }\sum_{k=0}^{M_1-1}\sum_{l=0}^{M_2-1}\tilde{H}(\omega_{1,k},\omega_{2,l})e^{j(\frac{2\pi}{M_1}km+\frac{2\pi}{M_2}ln)},
\end{align}
which can be implemented by using 2-D IFFT very fast, with complexity in the order of only $\mathcal{O}(M_1M_2(\log_2M_1+\log_2M_2))$. Finally, 
\begin{align}\label{eq:hfinal}
    h(m,n)=e^{-j(m+n)\pi}\tilde{h}(m,n),
\end{align}
for $m=0,1,...,\frac{N_x}{2}-1$, and $n=0,1,...,N_y-1$.

\subsection{Algorithm Summary}
Now we can summarize the proposed fast algorithm for RIS beam pattern design as in Algorithm-\ref{alg:algorithm1}: 

\begin{algorithm}[h]
    \caption{Proposed fast RIS beam pattern design}
    \label{alg:algorithm1}
    \KwIn{RIS related: $N_x, N_y, d=\frac{\lambda}{2}$; Input signal related: $\{\theta_{in,i}^{azi},\theta_{in,i}^{ele}\}_{i=1}^I$; Desired beam pattern related: $|\hat{g}(\theta^{azi},\theta^{ele})|, M_1, M_2$.
}
    \KwOut{$v(n_x,n_y), n_x=0,1,\cdots,N_x-1, n_y=0,1,\cdots,N_y-1$.}
    \BlankLine
    Construct desired beam pattern in transform domain: 
    \ForEach{$k=0,1,\cdots,M_1-1,l=0,1,\cdots,M_2-1$}{
            $\omega_{k,l}=\sqrt{\omega_{1,k}^2+\omega_{2,l}^2}$\;
            \eIf{$\omega_{k,l}\leq \pi$}{
            
    $\hat{H}(\omega_{1,k},\omega_{2,l})=|\hat{g}(\theta_{k,l}^{azi},\theta_{k,l}^{ele})|$, where $\theta_{k,l}^{azi}=\angle (\omega_{1,k}+j\omega_{2,l})$, $\theta_{k,l}^{ele}=\arcsin{\left(\frac{ \omega_{k,l}}{\pi}\right)}$;
     }
    {$\hat{H}(\omega_{1,k},\omega_{2,l})=0$;}
    Obtain $\tilde{H}(\omega_{1,k},\omega_{2,l})$ by \eqref{eq:tildeH} where $L_1$ and $L_2$ are replaced by $N_x$ and $N_y$;
   }

Obtain $\tilde{h}(n_x,n_y),n_x=0,1,\cdots,M_1-1,n_y=0,1,\cdots,M_2-1$, by performing 2-D IFFT on $\tilde{H}(\omega_{1,k},\omega_{2,l})$;

Obtain $h(n_x,n_y)$, $n_x=0,1,\cdots,\frac{N_x}{2}-1$, $n_y=0,1,\cdots,N_y-1$, by \eqref{eq:hfinal}. 

Obtain the other half of $h(n_x,n_y)$ by \eqref{eq:conj}.

Obtain $v(n_x,n_y)$ by \eqref{eq:vfinal}.
\end{algorithm}

Remarks: 1) In line 4 of Algorithm-\ref{alg:algorithm1}, $j=\sqrt{-1}$ and $\angle z $ denotes the phase angle of a complex number $z$ in the interval $[0,2\pi)$; 2) In the above procedure, the filter design stage will introduce Gibbs effect, which can be alleviated by smoothing the abrupt changing in the desired magnitude response; 3) In practice, quantization on the amplitude and phase may be needed. This can be performed directly on $v(n_x,n_y)$. It can be seen in the simulation part that a few bits 
quantization on both amplitude and phase will be sufficient to keep the performance. 

\section{Simulation Results}\label{sec:sim}
This section provides simulation results of the proposed algorithm. Comparisons with baseline method in \cite{WGGR21} are also given. The simulation setup is as follows. Two beam spots are desired. One is a $(\frac{\pi}{3}\times \frac{\pi}{6})$-sized rectangular centered at $(\frac{\pi}{2},\frac{\pi}{4})$ with magnitude 1. The other is a circular of diameter $\frac{\pi}{6}$ centered at $(\frac{3\pi}{2},\frac{\pi}{4})$ with magnitude 0.5. 

\begin{figure}[pbth]
    \centering
    \includegraphics[width=0.96\linewidth,draft=false]{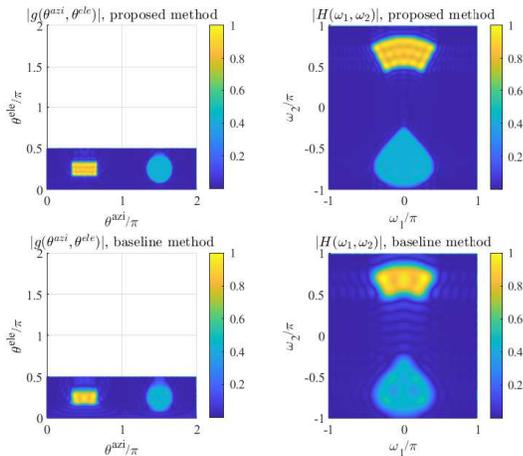}
    \setlength{\abovecaptionskip}{-10pt} 
    \caption{The designed beam patterns of the proposed method and the baseline method.}
    \label{fig:designed}
\end{figure}

\begin{figure}[pbth]
    \setlength{\abovecaptionskip}{0pt} 
\centering
    \includegraphics[width=0.95\linewidth,draft=false]{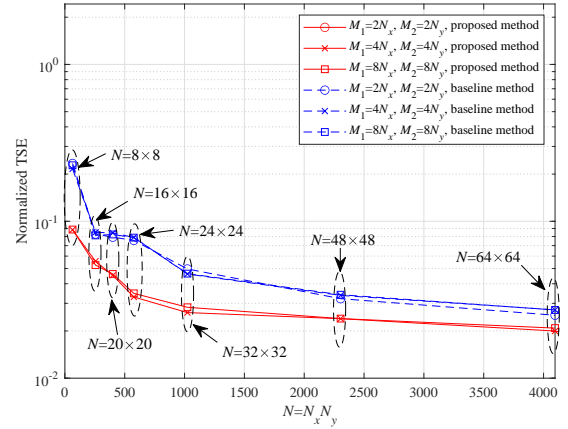}
    \caption{The normalized TSE performance of the proposed method and the baseline method.}
    \label{fig:TSE}
    
\end{figure}

Fig. \ref{fig:designed} shows the designed beam patterns using our proposed method and the baseline method. The following parameters are assumed: $N_x=N_y=32$, $M_1=4N_x$, $M_2=4N_y$. It can be seen that our proposed method provides higher quality of beam patterns in both domains. To quantify the performance, Fig. \ref{fig:TSE} plots the normalized TSE performance of the designing methods with various parameter configurations, where $N=N_x\times N_y$. It can be seen that the TSE is more relevant to the number of reflecting units than the size of sampling grid $M_1\times M_2$. In any case, our proposed method always outperform the baseline method. 

\begin{figure}[pbth]
\setlength{\abovecaptionskip}{0pt}
\setlength{\belowcaptionskip}{-1cm} 

    \centering
    \includegraphics[width=0.95\linewidth,draft=false]{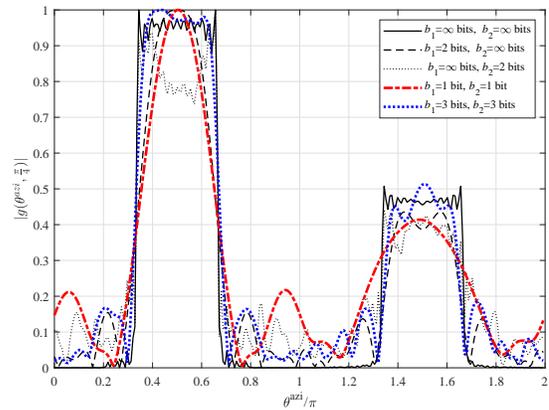}
    \caption{Beam patterns in the cross section with different quantization resolutions.}
    \label{fig:magnitude}
\end{figure}

Fig. \ref{fig:magnitude} shows beam patterns along the 2-D cross section of $\theta^{ele}=\frac{\pi}{4}$ with different quantization resolutions. Here, $b_1$ and $b_2$ denote the number of quantization bits for amplitude and phase of the reflection coefficients, respectively. It can be observed that compared with the beam pattern without quantization, a few bits of quantization will preserve the performance to a large extent.  

\section{Conclusions}\label{sec:conclusion}
This paper presented a fast solution for arbitrary beam pattern design in RIS-assisted terahertz wireless communication. The proposed method links beam pattern design with 2-D FIR filter design. It shows improved performance compared with existing method. With this convenient interpretation, the proposed method may facilitate many exciting works such as fast wireless positioning\footnote{A simulation demo is associated with this paper.} and over-the-air computation.

\hfill

\end{document}